\documentclass[11pt,prd]{revtex4-1}
\usepackage{graphicx, amsmath, amssymb, url, color}
\def \kpiv{k_*}
\def \primsca{\mathcal{P}_s}

\def \figurewidth{0.65\textwidth}
\def \halffigurewidth{0.45\textwidth}

\def \vecx{\mathbf{x}}
\def \veck{\mathbf{k}}
\def \vecn{\mathbf{n}}
\def \vel#1{\upsilon_{#1}}

\begin{document}
\title{A Cosmology Forecast Toolkit -- CosmoLib}
\author{Zhiqi Huang}
\affiliation{  CEA, Institut de Physique Th{\'e}orique, 91191 Gif-sur-Yvette c{\'e}dex, France\\ CNRS, URA 2306, F-91191 Gif-sur-Yvette, France}
\date{\today}
\begin{abstract}
The package CosmoLib is a combination of a cosmological Boltzmann code and a simulation toolkit to forecast the constraints on cosmological parameters from future observations. In this paper we describe the released linear-order part of the package. We discuss the stability and performance of the Boltzmann code. This is written in Newtonian gauge and including dark energy perturbations. In CosmoLib the integrator that computes the CMB angular power spectrum is optimized for a $\ell$-by-$\ell$ brute-force integration, which is useful for studying inflationary models predicting sharp features in the primordial power spectrum of metric fluctuations. As an application, CosmoLib is used to study the axion monodromy inflation model that predicts cosine oscillations in the primordial power spectrum. In contrast to the previous studies by  Aich {\it et al} and Meerburg {\it et al}, we found no detection or hint of the osicllations. We pointed out that the CAMB code modified by Aich {\it et al} does not have sufficient numerical accuracy. CosmoLib and its documentation are available at \url{http://www.cita.utoronto.ca/~zqhuang/CosmoLib}.
\end{abstract}
\maketitle

\section{Introduction}

The hot big bang model and the cosmological perturbation theory, where the physical metric is perturbed around the spatially homogeneous and isotropic Friedmann-Robertson-Walker (FRW) metric \cite{Peebles:1980, Peacock:1999, Mukhanov:2005, Weinberg:2008a}, have led to a remarkable success in interpreting the plethora of observational data of the last two decades \cite{Riess/etal:2011, Amanullah/etal:2010, Sullivan/etal:2011, Reid/etal:2010, Percival/etal:2010, Komatsu/etal:2011}. Observations of the temperature anisotropy in the Cosmic Microwave Background (CMB) have been playing an essential role in building the standard cosmological model and measuring its parameters \cite{Komatsu/etal:2011, Fixsen/etal:1996}. In order to maximize the usage of the observational data, one would like to compute the theoretical prediction on the CMB anisotropy for a given model as accurately as possible, with tolerable time consumption. Computation tools developed over the years such as CMBFAST \cite{Seljak/Zaldarriaga:1996}, CAMB \cite{Lewis/etal:2000}, CMBEASY \cite{Doran:2005}  and  CLASS \cite{Lesgourgues:2011, Blas/etal:2011} are capable of computing a CMB angular power spectrum to percent-level accuracy within a few seconds on a modern desktop personal computer. 

The crucial technique used in all the fast CMB codes to date is the line-of-sight integration approach \cite{Seljak/Zaldarriaga:1996, Hu/White:1997, Hu/etal:1998} and an assumption that the primordial power spectrum of metric perturbations is smooth. (Here for readability we focus on the comoving curvature perturbations and temperature anisotropies,  although the same arguments can be as well applied to the tensor perturbations and CMB polarization.) A CMB code first computes the radiation transfer function $\Delta_\ell^k$ by solving the linear-order Boltzmann equations and using the line-of-sight integration method, then convolves $|\Delta_\ell^k|^2$ with the primordial power spectrum $\mathcal{P}(k)$ to obtain the CMB angular power spectrum $C_\ell$. The smoothness assumption allow us to compute only a few tens of multipoles spanning from $\ell_{\min} = 2$ to $\ell_{\max}\sim$ a few thousands and interpolate the remaining $C_\ell$'s. Furthermore, since $\mathcal{P}(k)$ is assumed to a smooth function, sparse sampling of the radiation transfer function has been implemented for the integration of each $C_\ell$.

A smooth primordial power spectrum is a prediction of the simplest single-field slow-roll inflation models \cite{Abbott/Wise:1984, Lucchin/Matarrese:1985, Lucchin/Matarrese:1985a, Lyth/Stewart:1992, Stewart/Lyth:1993}. However, local signatures in the primordial power spectrum that makes it deviate from smoothness can arise in various alternative models, for instance, when the inflaton potential has sharp features \cite{Starobinsky:1992,Chen/etal:2007}, when there is  a transition between different stages in the inflaton evolution  \cite{Contaldi/etal:2003,Cline/etal:2003}, when more than one field is present \cite{Polarski/Starobinsky:1992,Langlois/Vernizzi:2005}, from particle production during inflation \cite{Barnaby/etal:2009a, Barnaby/Huang:2009}, modulated preheating ~\cite{Chambers/Rajantie:2008,Bond/etal:2009}, or, more recently, in models motivated by monodromy in the extra dimensions \cite{Silverstein/Westphal:2008} (see also \cite{Bean/etal:2008}). These features represent an important window on new physics because they are often related to UV scale phenomena inaccessible to experiments in the laboratory. For these models, the CMB angular power spectrum is not necessarily smooth, and therefore needs to be computed at each multipole without interpolation. This increases the computing time by a factor of a few tens. Moreover, for the numerical-integration of each $C_\ell$, the sampling frequency in the wavenumber $k$ often needs to be increased, again, by a factor of a few tens. The required sampling frequency in $k$ is model-dependent. It is determined by the larger between the minimum width of the features in the primordial power spectrum and the minimum width of the oscillations in the radiation transfer function. 

To keep track of the features in the primordial power spectrum, one can modify standard CMB codes by naively doing an $\ell$-by-$\ell$ brute-force calculation with increased integration sampling frequency in $k$. However,  in the case where the features in the primordial power spectrum are really sharp ($\delta\ln k \lesssim 0.01$), this naive modification increases the computing time by a factor of $\sim 10^3$ (a few tens in $\ell$ sampling times a few tens in $k$ sampling). Moreover, the  memory that is required to store all the transfer functions and tables of spherical Bessel functions can be too large for most desktop personal computers. One of the purposes of this paper is to introduce a more optimized algorithm to treat these problems. In fact, apart from increasing the sampling frequency, that cannot be avoided, all the other problems can be significantly alleviated by using the recurrence relation of spherical Bessel functions. An optimized algorithm, which we detail in Section~\ref{sec:cmb}, is $\lesssim 10^2$ times slower than the standard algorithm for the smooth-$\mathcal{P}(k)$ case. This new algorithm has been implemented in the CosmoLib package, a self-contained package that we developed to compute cosmological perturbations, CMB angular power spectra, and the forecast constraints on cosmological parameters from future cosmological surveys using Fisher matrix analysis and Monte Carlo Markov Chain (MCMC) calculation. In particular, the cosmological surveys that we consider are CMB, large scale structure (LSS) and supernovae (SN). 

In addition to the enhanced CMB integrator, CosmoLib has a few other features that are complementary to the other publicly available Boltzmann/CMB/MCMC codes. For instance, the MCMC engine in CosmoLib has a modified rejection rule that allows the proposal density  (the probability of random-walking to a new point in the parameter space) to periodically depend some parameters. This is useful when one considers a likelihood that depends on some periodic parameter. This happens, for instance, in the context of inflation from axion monodromy \cite{Silverstein/Westphal:2008,Chen/etal:2008,McAllister/etal:2010, Flauger/etal:2010}, where the oscillations in the predicted power spectrum depend on a free phase. Moreover, CosmoLib treats the dark energy equation of state (EOS) $w(a)$ and the primordial scalar and tensor power spectra $\mathcal{P}_s(k)$ and $\mathcal{P}_t(k)$, as free functions, which can be either chosen from a list of build-in models or defined by the user. This makes CosmoLib a convenient tool to study non-standard parametrizations of dark energy EOS and primordial power spectra. Finally, CosmoLib is written in Newtonian gauge (also called Poisson gauge) \cite{Mukhanov/etal:1992,Ma/Bertschinger:1995, Bertschinger:1996}, while many other codes are mainly developed in synchronous gauge (see e.g.~\cite{Peacock:1999}).  This is a plus-and-minus point. We found that our Newtonian-gauge Boltzmann code is slightly slower than the codes written in synchronous gauge. However, many theoretical works in the literature have derived equations in Newtonian gauge. For instance, second-order Boltzmann equations have been derived in this gauge \cite{Bruni/etal:1997, Bartolo/etal:2006, Bartolo/etal:2007, Nakamura:2007, Enqvist/etal:2007, Nakamura:2009, Malik/Wands:2009, Boubekeur/etal:2009, Pitrou:2009, Nitta/etal:2009, Senatore/etal:2009, Nakamura:2010, Beneke/Fidler:2010, Bernardeau/etal:2011}. Implementing these equations in a code already in Newtonian gauge would be much easier. To conclude the discussion, we list the differences between CosmoLib and other publicly available CMB codes in Table~\ref{tbl:codes}. 
\begin{table}
\caption{Comparison between CMB Codes \footnote{Here we do not include CMBFast, which is no longer supported by its authors or available for download.} \label{tbl:codes}}
\begin{tabular}{cccccc}
\hline
\hline
 & CAMB & CLASS & CMBEASY & CMBquick & CosmoLib \footnote{This refers to CosmoLib Version 0.2.} \\
Language & F90 & C & C++ & Mathematica & F90\footnote{CosmoLib is a mixture of Fortran and C codes. The main part is written in Fortran.}  \\
gauge \footnote{syn.: synchronous gauge; Newt.: Newtonian gauge; gauge-inv.: gauge-invariant variables.} & syn. & syn./Newt. \footnote{Newtonian gauge is implemented in CLASS version 1.3.} & syn./gauge-inv. & Newt. & Newt. \\
open/close universe & Yes & No & No & No & No  \\
massive neutrinos & Yes & Yes & Yes & Yes & No \\
tensor perturb. & Yes & Yes & Yes & Yes & Yes \\
CDM isocurvature mode & Yes & Yes & Yes & Yes & Yes \\
dark energy perturb. & Yes & Yes & Yes & No & Yes \\
nonzero $c^2_{s,b}$ &  Yes & Yes & Yes & No & Yes \\
dark energy EOS. & constant\ &\ $w_0+w_a(1-a)$\ &\ arbitrary & -1 & arbitrary \\
non-smooth primordial power & No & No & No & No & Yes \\
MCMC driver & Yes & No & Yes & No & Yes \\
periodic proposal density & No & No & No & No & Yes \\
data simulation & No & No & No & No & Yes \\
second-order perturb. \footnote{A second-order perturbation code is used to study the CMB non-Gaussianity. } & No & No & No & Yes  & No \footnote{The second-order part of CosmoLib is not released with this paper.} \\
\hline
\end{tabular}
\end{table}

As an application, CosmoLib is used to study the ``hints'' of cosine osicllations in the primordial power spectrum that was recently found in Refs.~\cite{Aich/etal:2011, Meerburg/etal:2012}. In an accompanying paper \cite{Huang/etal:2012}, CosmoLib is applied to forecast the constraining power of future CMB and galaxy survey data on the primordial power spectrum from inflation, with an emphasis on models generating features in the power spectrum. 

This paper is organized as follows. In Section~\ref{sec:perturb} we introduce the Boltzmann code in Newtonian gauge and discuss its stability and performance. Section~\ref{sec:cmb} details the algorithm used in the enhanced CMB integrator. In Section~\ref{sec:forecast} we introduce the forecast technique and parameter estimation methods. Section~\ref{sec:conclusion} concludes.

Throughout this paper, unless otherwise specified, repeated indices are summed over. Greek indices run from 0 to 3. Latin indices run from 1 to 3, that is only over spatial dimensions. We use natural units $c=\hbar = 1$ and the reduced Planck Mass $M_p\equiv 1/\sqrt{8\pi G_N} = 2.43 \times 10^{18} {\rm GeV}$. 

\section{CosmoLib in Newtonian Gauge \label{sec:perturb}}

\subsection{The Background Solutions}

Let us start discussing the background solutions. We consider a flat FRW metric
$ds^2 = a^2(\tau)(-d\tau^2 + dx^idx^i) $, where $a$ is the scale factor and $\tau$ is the conformal time. The normalization of $a$ is arbitrary. We normalize it such that $a=1$ today. CosmoLib uses the e-fold number $N \equiv \ln a$ as the time variable. The physical Hubble expansion rate is defined as $H \equiv \frac{da/d\tau}{a^2}$. Its present value is denoted by $H_0 \equiv 100 h {\,\rm km\,} {\rm s}^{-1} {\,\rm Mpc}^{-1}$. 

We assume a universe with cold dark matter (labeled with a subscript $c$), dark energy (labeled with a subscript $\Lambda$), baryons (labeled with a subscriber $b$), radiation (labeled with a subscript $\gamma$), and 3 species of massless neutrinos (labeled with a subscript $\nu$). For a component $X$ ($X= b, c, \gamma, \nu, \Lambda$) the background density is denoted as $\rho_X$, and the background pressure $p_X$. The present-day fractional energy density is written as $\Omega_{X0}$ . Dark energy is assumed to be a perfect fluid with known equation of state (pressure to density ratio) $w(a)$ and a constant sound speed $c_{s,\Lambda}^2$ in its rest frame. The users can either choose $w(a)$ from a list of build-in models or define their own $w(a)$ functions. The build-in models of $w(a)$ include the cosmological constant model $w(a)=-1$ \cite{Einstein:1917}, a constant EOS $w(a)=w_0$, a linear function $w(a)=w_0+w_a(1-a)$  \cite{Chevallier/Polarski:2001,Linder:2003}, and a general three-parameter parametrization for the minimally coupled quintessence/phantom models \cite{Huang/etal:2011}.

For a given $w(a)$ the background solutions are
\begin{equation}
\begin{split}
a &= e^N, \\
\rho_c &= 3 H_0^2M_p^2 \Omega_{c0} a^{-3}\;,\; p_c = 0\;, \\
\rho_b &= 3 H_0^2M_p^2 \Omega_{b0} a^{-3}\;,\; p_b = 0\;, \\
\rho_\gamma &= 3 H_0^2M_p^2 \Omega_{\gamma 0} a^{-4}\;,\; p_\gamma = \frac{1}{3} \rho_\gamma \; , \\
\rho_\nu &= 3 H_0^2M_p^2 \Omega_{\nu 0} a^{-4}\;,\; p_\nu = \frac{1}{3} \rho_\nu\;, \\
\rho_\Lambda &= 3 H_0^2M_p^2 \Omega_{\Lambda 0} a^{-3} \exp{\left[3\int_N^0 w(a) dN\right]}\;,\; p_\Lambda = w(a) \rho_\Lambda\; ,\\
H &= \frac{1}{M_p}\sqrt{\frac{\rho_c+\rho_b+\rho_\gamma +\rho_\nu+\rho_\Lambda }{3}}\; . 
\end{split}
\end{equation}
We will also use the derived quantities $\Omega_X(a) \equiv \rho_X/(3H^2M_p^2)$ ($X= b, c, \gamma, \nu, \Lambda$), $R\equiv (3\rho_b)/(4\rho_\gamma)$, and 
\begin{equation}
\epsilon =  -\frac{d\ln H}{dN} = \frac{3}{2}\left[1+\frac{p_\Lambda + p_\gamma + p_\nu}{\rho_c+\rho_b+\rho_\Lambda+\rho_\gamma + \rho_\nu}\right]\; .
\end{equation}
The conformal time $\tau$ can be related to the scale factor $a=e^N$ via 
\begin{equation}
  \tau = \int_0^a\frac{da}{Ha^2}.
\end{equation}

The electron number density $n_e(a)$ is obtained using RecFast version 1.5 \cite{Seager/etal:1999, Wong/etal:2008}, which has been incorporated into CosmoLib. We denote the differential optical depth (increment of optical depth per $dN$) as
\begin{equation}
  \kappa_N \equiv \frac{d\kappa}{dN} = \frac{n_e \sigma_T}{H}\; ,
\end{equation}
where $\sigma_T = 6.652\times 10^{-25}{\rm cm}^2$ is the Thomson scattering cross section. The baryon sound speed $c_{s,b}^2(a)$ is obtained by solving the differential equations~(68-69) in Ref.~\cite{Ma/Bertschinger:1995}. 

With these background solutions in hand, now we can write down the governing equations for scalar perturbations.

\subsection{Scalar Perturbations}

The metric in the (generalized) Newtonian gauge can be written as
\begin{equation}
ds^2 = a^2(\tau) \left\{-(1 + 2\Phi)d\tau^2 + \omega_i dx^id\tau + \left[(1-2\Psi)\delta_{ij} + h_{ij}\right]dx^idx^j \right\} \;,
\end{equation}
where $\partial_i\omega_i = 0$, $h_{ii}=0$ and $\partial_i h_{ij} = 0$. The vector perturbation $\omega_i$ decays in an expanding universe and hence it is set to zero in CosmoLib. The tensor perturbation $h_{ij}$ is gauge-invariant and its governing equations are identical in all gauges. Thus, we will only focus on the scalar perturbation equations that in CosmoLib differ from those in many other Boltzmann codes.

The linear-order relative density perturbation of $X$ is denoted by $\delta_X \equiv \delta\rho_X/\rho_X$, and the linear-order velocity $\vel{X}$. Unless otherwise specified, $\delta_X$ and $\vel{X}$ are all defined in Fourier space, that are functions of $\tau$ and the wave vector $\veck$. 

The radiation relative temperature fluctuation  $\Delta T/T$ from direction $\vecn$ seen by an observer at position $\vecx$  is expanded as \cite{Hu/White:1997}
\begin{equation}
\frac{\Delta T}{T} \left(\vecx, \vecn, \tau \right)= \int \frac{d^3\veck}{(2\pi)^3}\sum_{\ell=0}^{\infty}\sum_{m=-2}^{2} \Theta_\gamma(\ell,m) (-i)^\ell \sqrt{\frac{\pi}{4(2\ell +1)}}Y_\ell^m(\vecn) e^{i\veck\cdot\vecx}\; ,
\end{equation}
where $Y_{\ell}^m$ are the spherical harmonic functions. Note that the moments $\Theta_\gamma(\ell,m)$ are functions of the wavenumber $\veck$ and the conformal time $\tau$. The energy density fluctuation and velocity of photons are related to the moments $\ell = 0$ and $\ell = 1$ via
\begin{eqnarray}
\delta_\gamma = \Theta_\gamma(0,0);\ \;
\vel{\gamma} = \frac{1}{4}\Theta_\gamma(1,0)\; .
\end{eqnarray}
The neutrino moments $\Theta_\nu(\ell,m)$ are defined in the same way, by replacing the subscript $\gamma$ with $\nu$.

For the polarization of radiation, the Stokes parameters $Q, U$ are expanded using the spin-$2$  harmonics  $_{\pm 2}Y_\ell^m$  \cite{Hu/White:1997}
\begin{equation}
\left(Q\pm i U\right)\left(\vecx, \vecn ,\tau\right) = \int \frac{d^3\veck}{(2\pi)^3}\sum_{\ell=0}^{\infty}\sum_{m=-2}^{2} \left[E(\ell,m) \pm B(\ell,m) \right] (-i)^\ell \sqrt{\frac{\pi}{4(2\ell +1)}}\left[_{\pm 2}Y_\ell^m(\vecn)\right] e^{i\veck\cdot\vecx}\; ,
\end{equation}
where $E(\ell, m)$ and $B(\ell, m)$ are functions of the wave vector $\veck$ and conformal time.

The linear-order Fourier modes are decoupled. The Fourier-space variables to be evolved are $\Psi$, $\Psi_N \equiv d\Psi/dN$, $\delta_b$, $\vel{b}$, $\delta_c$, $\vel{c}$, $\delta_\Lambda$, $\theta_\Lambda\equiv (1+w)\vel{\Lambda}$, $\Theta_\gamma(\ell, 0)$ ($\ell$ = $0$, $1$, $2$, ..., $\ell_{\max, \gamma}$), $\Theta_\nu(\ell, 0)$ ($\ell$ = $0$, $1$, $2$, ..., $\ell_{\max, \nu}$), $E(\ell, 0)$  ($\ell$ = $2$, ..., $\ell_{\max, E}$). The truncations $\ell_{\max, \gamma}$, $\ell_{\max, \nu}$ and $\ell_{\max, E}$ are adjustable integers. In CosmoLib their default values are taken to be $14$, $12$, $14$, respectively. Without loss of generality we choose the azimuthal direction (the $z$-axis direction that is used to define $Y_{\ell,m}(\vecn)$) to be parallel to  $\veck$. 

The gravitational potential $\Phi$ can be obtained from the Einstein equations \cite{Bond/Efstathiou:1984, Ma/Bertschinger:1995, Hu/White:1997}
\begin{equation}
\Phi = \Psi - \frac{3}{5k_H^2} \left[\Omega_\gamma \Theta_\gamma(2,0) + \Omega_\nu \Theta_\nu(2,0)\right]\; ,
\end{equation}
where we have introduced the reduced wavenumber 
\begin{equation}
k_H \equiv \frac{k}{aH}\; .
\end{equation}
Note that $k_H, \Omega_\gamma, \Omega_\nu$ are functions of time. We do not treat $\Phi$ as an independent-variable. Instead we view it as a function of the variables $\Psi$, $\Theta_\gamma(2,0)$ and $\Theta_\nu(2,0)$.

The close set of first-order differential equations including all the truncation schemes is:
{\allowdisplaybreaks
\begin{align}
\frac{d\Psi}{dN} &= \Psi_N\; ,\\
 \frac{d\delta_c}{dN} &= - k_H \vel{c} + 3 \Psi_N \;,\\
 \frac{d \vel{c}}{dN} &= - \vel{c} + k_H \Phi \; , \\
\frac{d\delta_b}{dN} &= - k_H \vel{b} + 3 \Psi_N \;, \\
\frac{d\vel{b}}{dN} &= - \vel{b} + k_H\left(\Phi + c_{s,b}^2\delta_b\right) - \frac{\kappa_N}{R} \left[\vel{b} - \frac{1}{4}\Theta_{\gamma}(1,0)\right]\;,\\
\frac{d\delta_\Lambda}{dN} &= - 3 \left(c_{s,\Lambda}^2 - w\right) \delta_\Lambda - 9\left[c_{\Lambda,s}^2 - \left(w-\frac{dw/dN}{3\left(1+w\right)}\right)\right] \frac{\theta_\Lambda}{k_H}  -  k_H \theta_\Lambda + 3 (1 + w) \Psi_N \;, \\
\frac{d\theta_\Lambda}{dN} &= 3 \left[w + c_{s,\Lambda}^2 - \left(w-\frac{dw/dN}{3\left(1+w\right)}\right) - \frac{1}{3}\right] \theta_\Lambda + k_H\left[c_{s,\Lambda}^2 \delta_\Lambda + (1+w) \Phi\right]\;,\\
\frac{d \Theta_\gamma(0,0)}{dN} &=  - \frac{1}{3} k_H \Theta_\gamma(1,0) + 4 \Psi_N\;,  \label{eq:hiestart}\\
\frac{d \Theta_\gamma(1, 0)}{dN} &= k_H\left[  \Theta_\gamma(0,0) - \frac{2}{5} \Theta_\gamma(2,0) + 4 \Phi\right] + \kappa_N \left[4\vel{b} - \Theta_\gamma(1,0)\right]\;,\\
\frac{d \Theta_\gamma(2, 0)}{dN} &= k_H\left[ \frac{2}{3} \Theta_\gamma(1,0) - \frac{3}{7} \Theta_\gamma(3,0) \right] - \kappa_N \left[\frac{9}{10}\Theta_\gamma(2,0)+\frac{\sqrt{6}}{10}E(2,0)\right]\;,\\
\frac{d \Theta_\gamma(\ell, 0)}{dN} &= k_H \left[\frac{\ell}{2\ell - 1} \Theta_\gamma(\ell - 1, 0) - \frac{\ell+1}{2\ell + 3} \Theta_\gamma(\ell+1, 0)\right] - \kappa_N \Theta_\gamma(\ell , 0)\; \;(2<\ell<\ell_{\max,\gamma})\;,\\
\frac{d\Theta_\gamma(\ell_{\max,\gamma}, 0)}{dN} &= \frac{2\ell_{\max,\gamma}+1}{2\ell_{\max,\gamma}-1} k_H\Theta_\gamma(\ell_{\max,\gamma}-1, 0) - \left(\kappa_N + \frac{\ell_{\max,\gamma}+1}{aH\tau}\right) \Theta_\gamma(\ell_{\max,\gamma},0)\;,\\
\frac{d \Theta_\nu(0,0)}{dN} &=  - \frac{1}{3} k_H \Theta_\nu(1,0) + 4 \Psi_N\;, \\
\frac{d \Theta_\nu(1, 0)}{dN} &= k_H\left[  \Theta_\nu(0,0) - \frac{2}{5} \Theta_\nu(2,0) + 4 \Phi\right]\;,\\
\frac{d \Theta_\nu(\ell, 0)}{dN} &= k_H \left[\frac{\ell}{2\ell - 1} \Theta_\nu(\ell - 1, 0) - \frac{\ell+1}{2\ell + 3} \Theta_\nu(\ell+1, 0)\right]\; \;(2\le\ell<\ell_{\max,\nu})\;,\\
\frac{d\Theta_\gamma(\ell_{\max,\nu}, 0)}{dN} &= \frac{2\ell_{\max,\nu}+1}{2\ell_{\max,\nu}-1} k_H\Theta_\nu(\ell_{\max,\nu}-1, 0) - \frac{\ell_{\max,\nu}+1}{aH\tau} \Theta_\nu(\ell_{\max,\nu},0)\;,\\
\frac{d E(2,0)}{dN} &= -k_H \frac{K_{3,0,2}}{7}E(3,0) - \kappa_N\left[\frac{2}{5} E(2,0) + \frac{\sqrt{6}}{10} \Theta_\gamma(2,0)\right]\; ,\\
\frac{dE(\ell,0)}{dN} &= k_H \left[\frac{K_{\ell,0,2}}{2\ell-1} E(\ell-1,0) - \frac{K_{\ell+1, 0, 2}}{2\ell+3} E(\ell+1,0)\right] - \kappa_N E(\ell,0) \;\;(2<\ell<\ell_{\max,E})\;, \\
\frac{dE(\ell_{\max,E},0)}{dN} &= \frac{2\ell_{\max,E}+1}{2\ell_{\max,E}-1}k_H E(\ell_{\max,E}-1,0) - \left(\kappa_N + \frac{\ell_{\max,E}+1}{aH\tau}\right)E(\ell_{\max,E},0)\;, \label{eq:hieend}\\
\frac{d\Psi_N}{dN} &= \frac{1}{2}\left\{(3c_{s,b}^2-1)\delta_b\Omega_b -\delta_c\Omega_c + \left[(3c_{s,\Lambda}^2-1) \delta_\Lambda + 9\left(c_{s,\Lambda}^2 - w+\frac{dw/dN}{3(1+w)}\right)\frac{\theta_\Lambda}{k_H}\right]\Omega_\Lambda\right\} - 2\Psi \nonumber \\ 
& - 2(1-\epsilon)\Phi   - \frac{k_H^2}{3}(2\Psi - \Phi) - (5-\epsilon)\Psi_N  + \frac{3}{5k_H^2}\left(\Omega_\gamma\frac{d\Theta_\gamma(2,0)}{dN} + \Omega_\nu\frac{d\Theta_\nu(2,0)}{dN}\right) \;. \label{eq:PsiNN}
\end{align}}
In the radiation and neutrino hierarchy equations (\ref{eq:hiestart}-\ref{eq:hieend}) we have used the Clebsch-Gordan coefficients $K_{l,m,s}$, which are defined as \cite{Hu/White:1997}
\begin{equation}
K_{\ell,m,s} = \left\{
\begin{array}{ll}
\frac{\sqrt{(\ell^2-m^2)(\ell^2-s^2)}}{\ell} & \text{, if\ } \ell \ge \max\{{|m|,|s|, 1}\}\; ; \\
0 & \text{, otherwise.} 
\end{array}
\right.
\end{equation}

These equations are already written in the form that can be directly implemented into a generic first-order ordinary-differential-equation (ODE) solver. For a derivation of these equations, see Refs.~\cite{Bond/Efstathiou:1984, Ma/Bertschinger:1995, Hu/White:1997, Fang/etal:2008}. (The change of time variable from $\tau$ to $N$ can be done straightforwardly using $d/d\tau = aH d/dN$.) The initial conditions can be found in Ref.~\cite{Ma/Bertschinger:1995}. For the tight-coupling approximation we follow Ref.~\cite{Doran:2005a}, where the obvious typos in Eqs.~(15-16) has been fixed. Since these treatments are identical to the original source, we do not repeat the discussion here. The interested readers are referred to these references for the governing equations and their derivation. 

CosmoLib allows the user-input $w(a)$ to be a phantom-crossing function, that is a function crossing the line $w=-1$. In this case we force $d\delta_\Lambda/dN$ and $d \theta_\Lambda/dN$ to be zero around the proximity of the phantom crossing. This is an approximation. Exact treatment requires input of at least one more degree of freedom \cite{Vikman:2005,Hu:2005, Caldwell/Doran:2005}, which cannot be implemented in a generic code. In Ref.~\cite{Fang/etal:2008} the reader can find an alternative treatment that works better for multiple scalar field models.

Equation~\eqref{eq:PsiNN} is the key equation that guarantees the numerical stability of the code (even for isocurvature initial conditions). It is obtained by subtracting the $ii$ components of the perturbed Einstein equations (pressure perturbations) from the $00$ component (density perturbations). This particular combination of the Einstein equations has been applied in the numerical code CMBquick \cite{Pitrou:2009, Pitrou/etal:2010}, which assumes that dark energy is a cosmological constant and ignores the baryon sound speed. Eq.~\eqref{eq:PsiNN} is a generalized version that includes dark energy perturbations and a nonzero baryon sound speed.

\begin{figure}
\includegraphics[width=\halffigurewidth]{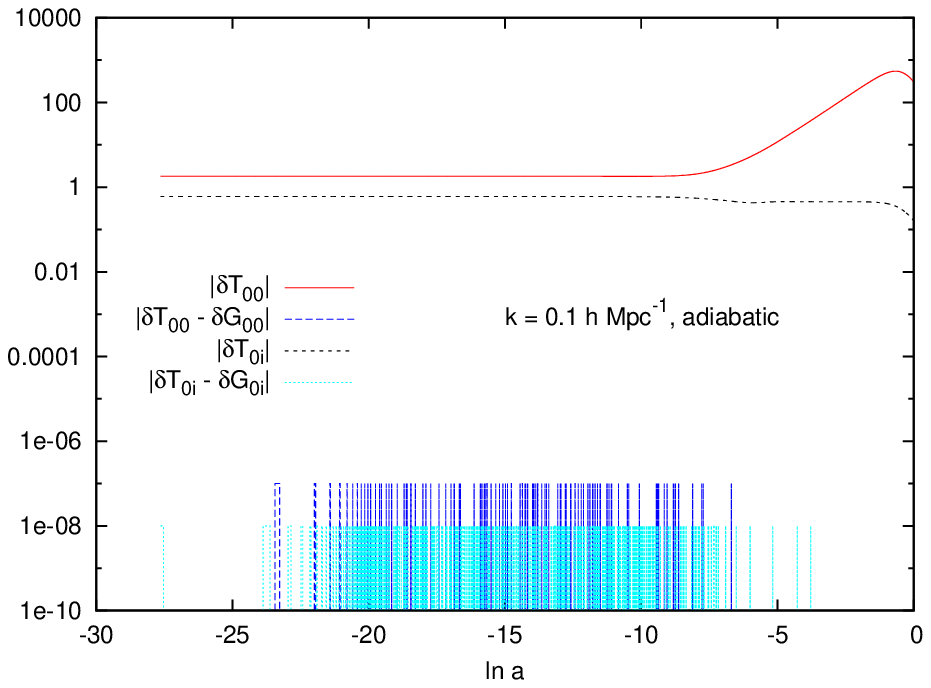}%
\includegraphics[width=\halffigurewidth]{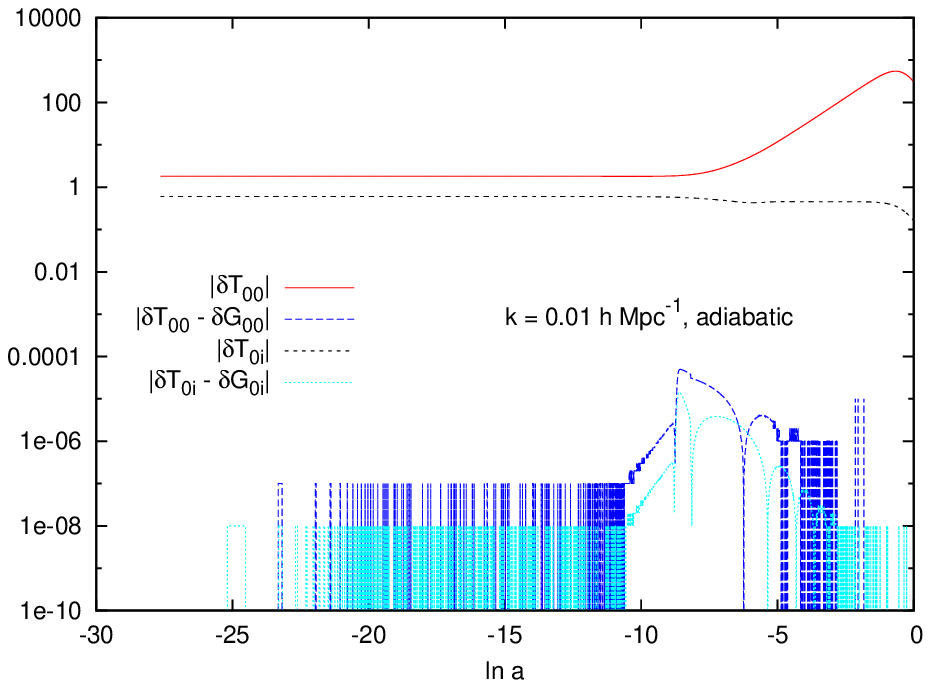}
\includegraphics[width=\halffigurewidth]{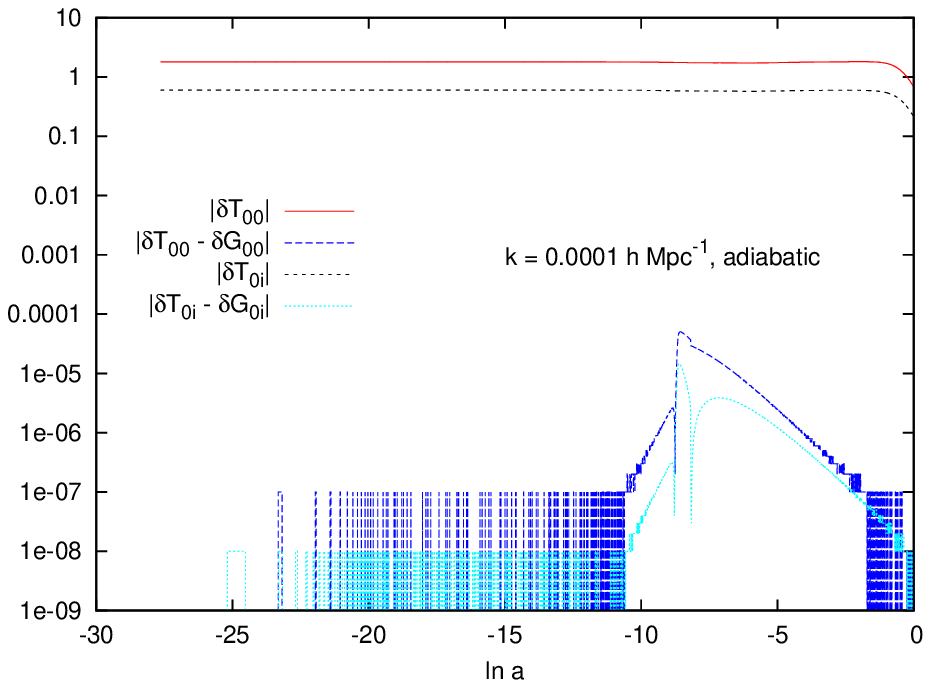}%
\includegraphics[width=\halffigurewidth]{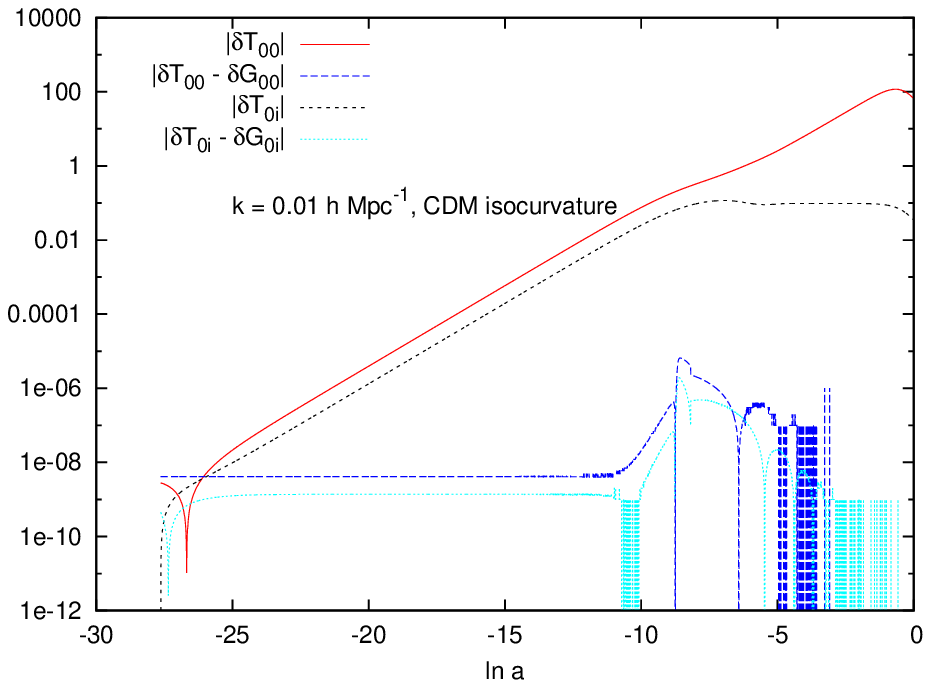}
\caption{Testing the energy constraint ($00$-component of the perturbed Einstein equation) and momentum constraint ($0i$-component of the perturbed Einstein equation).  The cosmological parameters used here are $\Omega_{b0} h^2 = 0.022$, $\Omega_{c0}h^2 = 0.1128$, $h = 0.72$. In the lower-right panel the CDM-isocurvature initial conditions are used, while in the other panels we have used adiabatic initial conditions. \label{fig:eins}}
\end{figure}

We can use the energy constraint ($00$ component of the perturbed Einstein equations, that is $\delta G_{00} = \delta T_{00}$) and the momentum constraint ($0i$-component of the perturbed Einstein equations, that is $\delta G_{0i} = \delta T_{0i}$) to estimate the numerical error of the code. As shown in shown in Figure~\ref{fig:eins}, the relative errors are $\lesssim 10^{-4}$ for a wide range of scales and different initial conditions.

\section{CMB Angular Power Spectra \label{sec:cmb}}

\subsection{Algorithm} 

Optionally CosmoLib can compute the CMB angular power spectrum for each multipole $\ell$ by brute force, i.e., without interpolation. The angular spectrum for the temperature anisotropies is given by 
\begin{equation}
C_\ell = \int |\Delta_\ell^k |^2 \primsca(k) d\ln k \;,  \label{eq:clint}
\end{equation}
where $\Delta_\ell^k$ is the temperature transfer function given by the line-of-sight integration
\begin{equation}
\Delta_\ell^k = \int_0^{\tau_0} S(k,\tau) j_\ell\left[k(\tau_0-\tau)\right] d\tau \; , \label{eq:trans}
\end{equation}
where $j_\ell$ is the spherical Bessel function and $\tau_0$ is $\tau$ at redshift zero. The source $S(k,\tau)$ can be computed from the perturbations $\Psi$, $\Phi$, $\delta_X$, $\vel{X}$ ($X=c$, $b$, $\Lambda$, $\gamma$, $\nu$), $\Theta_\gamma(2, 0)$ and $\Theta_\nu(2,0)$ \cite{Seljak/Zaldarriaga:1996, Hu/White:1997}. In Ref.~\cite{Hu/White:1997} the line-of-sight integration involves the functions $j_\ell$, $j'_\ell$ and $j_\ell''$. As shown in Ref.~\cite{Seljak/Zaldarriaga:1996}, however, the dependence on $j_\ell'$ and $j_\ell''$ can be eliminated by integrating by part. (We have corrected the obvious typos in eq.~(12b) in Ref.~\cite{Seljak/Zaldarriaga:1996}.)

\begin{figure}
  \includegraphics[width=\figurewidth]{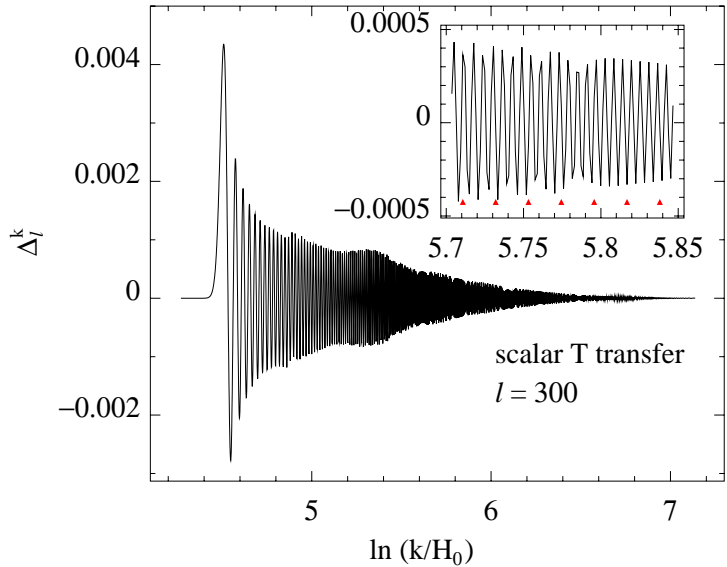}
  \caption{The temperature transfer function  $\Delta_\ell^k$ for a fixed $\ell = 300$.  A typical sampling scheme is shown by the red solid triangles in the upper-right panel, which zooms-in part of the figure. \label{fig:trans}}
\end{figure}

Since $\Delta_\ell^k$ is evaluated numerically and it typically oscillates quickly, its sampling is time consuming. Indeed, in modern fast CMB codes -- such as CAMB, CLASS, CMBEASY -- the integral~\eqref{eq:clint} is computed by sampling $\Delta_\ell^k$ using a step size in $k$ that can be typically much larger than the oscillation period in  $\Delta_\ell^k$. For instance, in Fig.~\ref{fig:trans} we show an example of $\Delta_\ell^k$ for a fixed $\ell = 300$. A typical sampling scheme is shown by the red solid triangles in the upper-right panel, which zooms-in part of the figure. According to Parseval's theorem, if $\primsca (\ln k)$ is a smooth function, such sparse sampling of $\Delta_\ell^k$ is enough. 

However, when $\primsca(k)$ has local sharp features, the minimum sampling distance should be determined by the larger between the typical relative width (the width measured in $\ln k$) of the oscillations in $\Delta_\ell^k$ and $\delta\ln k$, the typical relative width of the features in $\primsca(k)$. The former is about $10^{-4}-10^{-3}$, while the latter is model-dependent. For instance, if our goal is to sample features with width $\delta \ln k \sim 10^{-3}$, the required sampling frequency is typically $\sim 20 $ to  $100$ times higher than that used for a smooth $\primsca(k)$. Furthermore, as we wish to compute the $C_\ell$'s  for each $\ell$ rather than interpolating it over few tens of $\ell$'s, the total time consumption will be again multiplied by a factor of $\sim 10 - 50$. The naively estimated total time consumption is hence $\sim 10^3$ times more than that in the smooth-$\primsca(k)$ case. A final complication is due to the fact that, if all the transfer functions and the precomputed $j_\ell(x)$ tables are to be stored,  one has also to face a memory barrier that cannot be easily bypassed. For these reasons, simply increasing the $\ell$ and $k$ resolution in standard codes such as CAMB, CLASS or CMBEASY, will not be  efficient enough for the purpose of scanning the whole parameter space.

The algorithm can be significantly improved, however, if we notice that the output $S(k,\tau)$ from the Boltzmann code is a 2D matrix in $k$-$\tau$ space. If  $j_\ell\left[k(\tau_0-\tau)\right]$ is also a precomputed 2D matrix with the same structure, the integration~\eqref{eq:trans} can be obtained by taking the inner product of the two matrices. Modern Fortran90 compilers can optimize such operation and make the computation much faster. The difficulty, however, is that the $j_\ell\left[k(\tau_0-\tau)\right]$ matrices for all $\ell$'s will occupy too much memory (can be up to a few tens of Giga bytes in the worst scenario). Our solution is then to only store the matrices for two neighboring $\ell$'s and update them using the recurrence relation of spherical Bessel functions. 

Let us describe our strategy. We first compute two neighboring $C_\ell$'s by brute force. Two matrices of spherical Bessel functions $j_{\ell+1}[k(\tau_0 -\tau)]$ and $j_\ell[k(\tau_0-\tau)]$ are stored in the memory for each $(k, \tau)$ indices. Then we compute $C_{\ell-1}$. To do that, we update the $j_{\ell+1}$ matrix to the $j_{\ell -1}$ matrix using the recurrence relation 
\begin{equation}
  j_{\ell -1}(x) = \frac{2\ell +1}{x} j_\ell(x) - j_{\ell +1} (x) \ .\label{eq:jlrecur}
\end{equation}
Again, using $j_\ell$ and $j_{\ell-1}$ we then calculate $j_{\ell -2}$ and hence $C_{\ell -2}$. This downward iteration is very stable for a few tens of steps, after which we need to recompute another couple of neighboring $C_\ell$'s and iterate downward  again. 

The initial neighboring $j_{\ell}$'s are calculated using precomputed 25-th order Chebyshev fitting formulas. (For the rapidly oscillating part at $x\gg l$, the phase and amplitude of oscillations are fitted using Chebyshev polynomials.) Chebyshev fitting is slightly slower than the cubic-spline fitting used in other publicly available CMB codes, but it is more memory-efficient and more accurate -- it has an accuracy of $\sim 10^{-8}$ -- and allows more downward iterative steps. Finally, note that the algorithm proposed here is more efficient both CPU-wise and memory-wise, enhancing the speed of $\ell$-by-$\ell$ computation of $C_\ell$'s by a factor of $\sim 10$ to $30$. 

For CMB lensing we use the power spectrum approach as described in Refs.~\cite{Seljak:1996, Zaldarriaga/Seljak:1998}. 

\subsection{Testing the Code}

The trivial comparison between CosmoLib and CAMB for smooth-$\primsca$ models can be found in the online documentation at \url{http://www.cita.utoronto.ca/~zqhuang/CosmoLib}. Here we focus on the enhanced CMB integrator that does not assume the smoothness in $\primsca(k)$. Since this feature is not available in other CMB codes, direct numerical comparison is not possible when there is very sharp features in $\primsca$. Thus, we need to study a model in which we have some theoretical insights. An ideal candidate is the axion monodromy inflation model, where the primordial power spectrum displays sinusoidal oscillations superimposed to a smooth power spectrum. It can be written as \cite{Flauger/etal:2010}
\begin{equation}
   \primsca(k) = A_s  \left(\frac{k}{\kpiv}\right)^{n_s-1} \left[1 + \delta n_s \cos \left(\frac{\ln (k/\kpiv)}{\delta\ln k} + \varphi  \right)\right]\; , \label{eq:monoPs}
\end{equation}
where $A_s$ and $n_s$ are the amplitude and tilt, respectively.
The parameter $\delta\ln k$ describes the width of the oscillations in $\primsca(k)$, while $\delta n_s$ gives their relative amplitude. The pivot scale $\kpiv$ is chosen to be $0.05{\rm Mpc}^{-1}$ in our computation.

\begin{figure}
\centering
\includegraphics[width=\figurewidth]{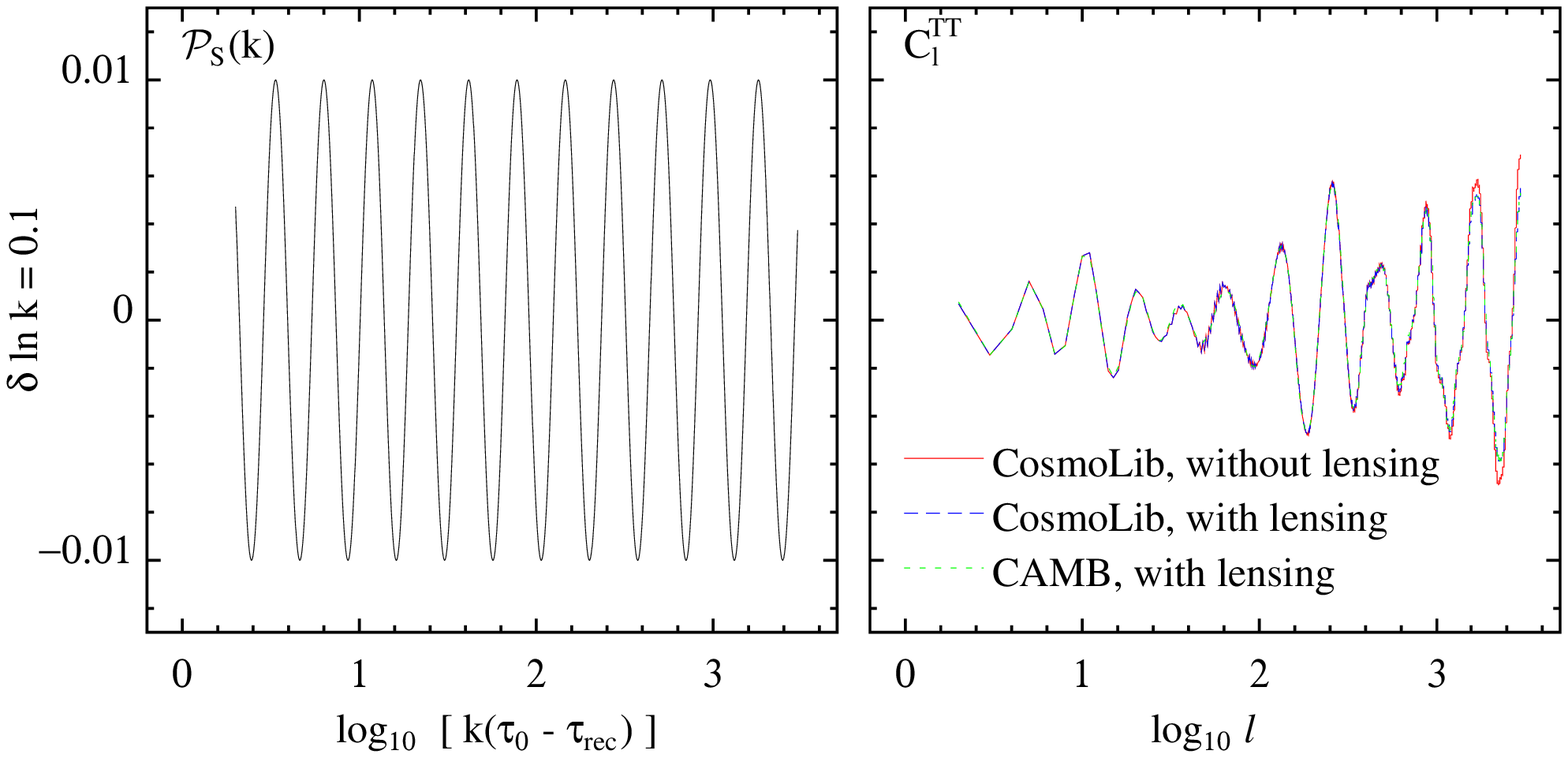}
\includegraphics[width=\figurewidth]{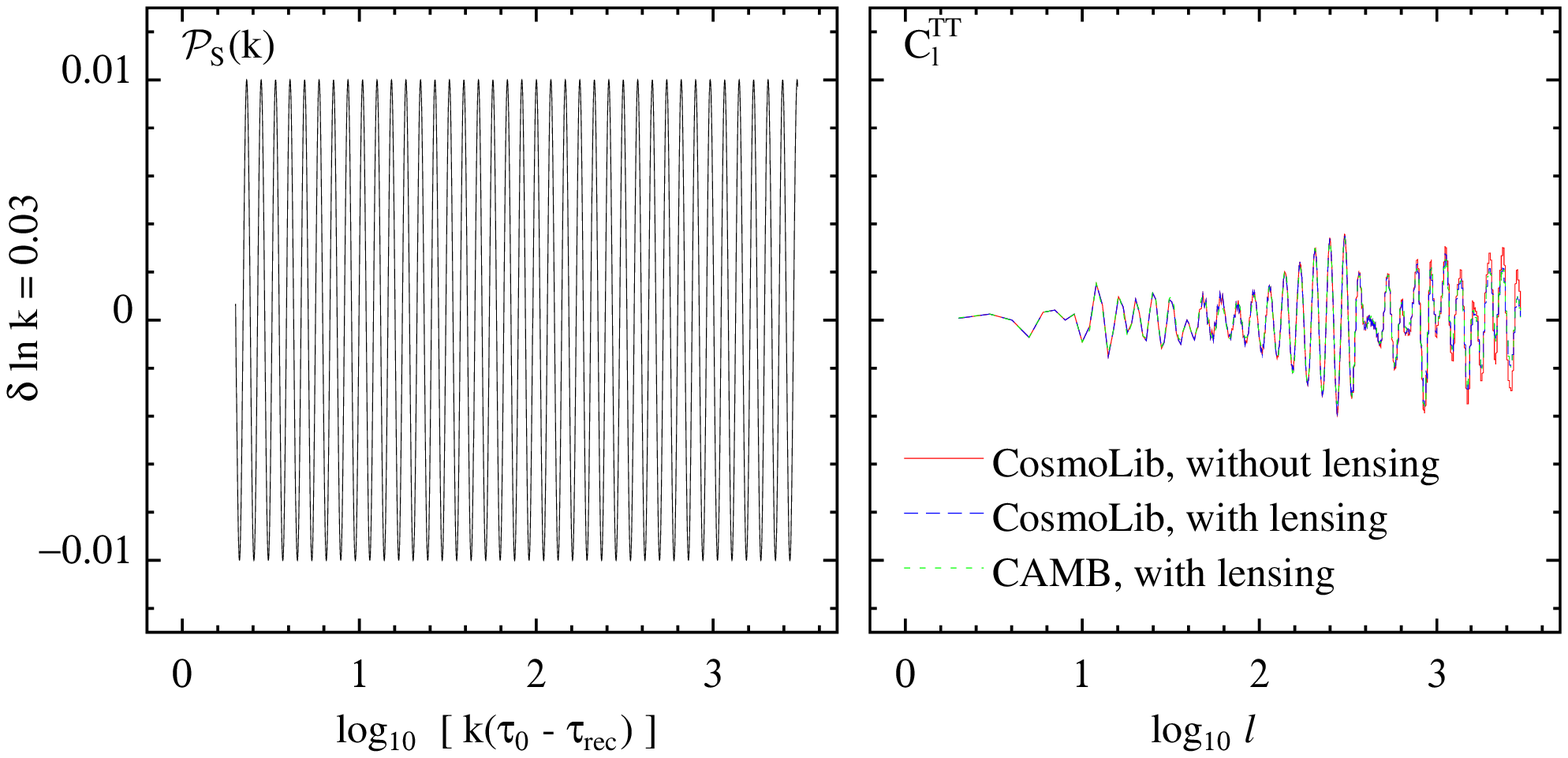}
\includegraphics[width=\figurewidth]{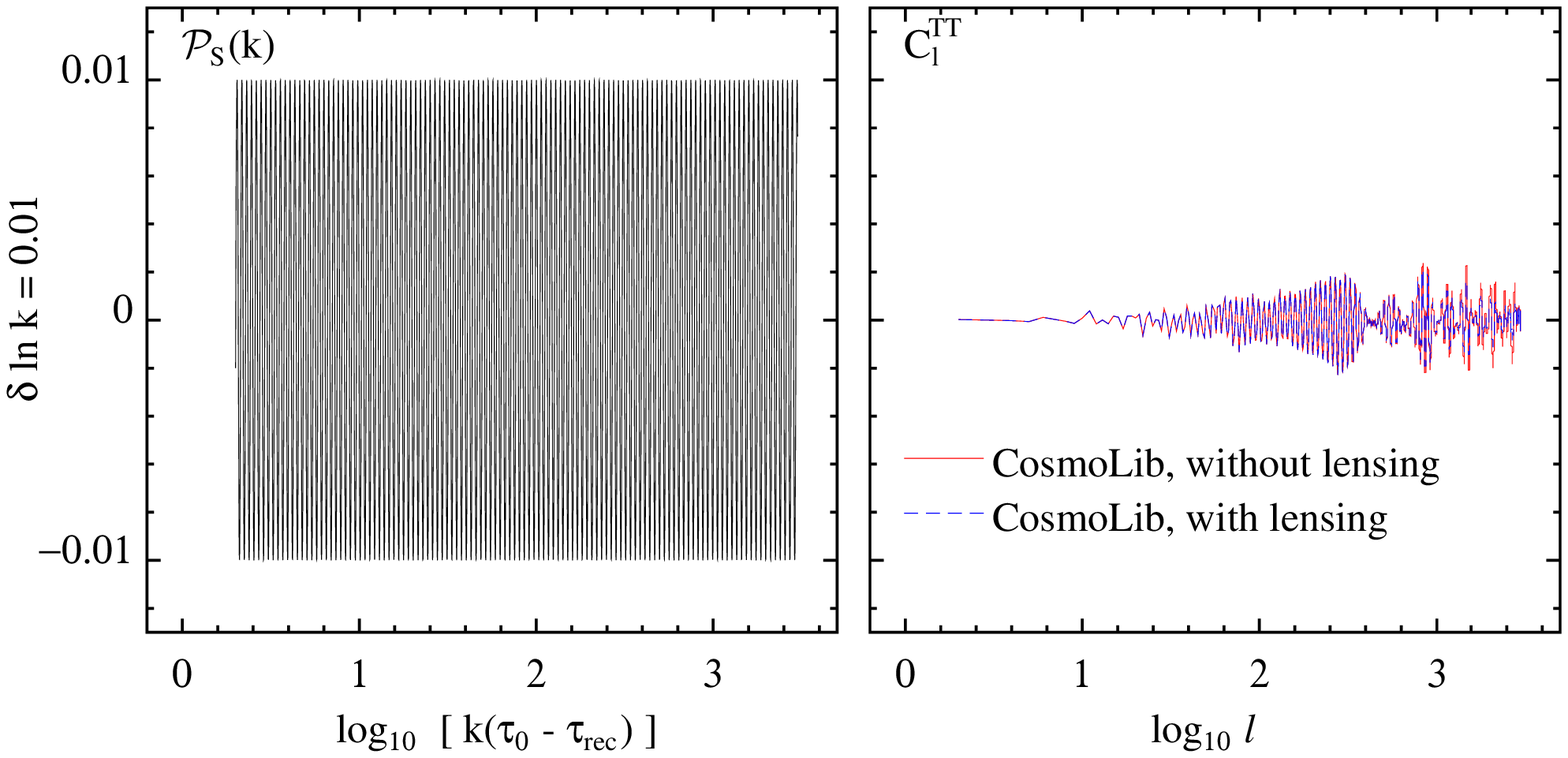}
\caption{The differences in $\ln \primsca$ (left panels) or $\ln C^{TT}_\ell$ (right panels) between a fiducial axion monodromy model with  $\ln \left(10^{10}A_s\right)=3.027$, $n_s = 0.975$, amplitude of cosine modulation $\delta n_s = 0.01$, phase $\varphi=0$  and a smooth power-law spectrum with the same  $A_s$ and $n_s$. For the top to bottom a series of $\delta\ln k = 0.1$, $0.03$, $0.01$ are used, respectively. The  $\tau_{\rm recomb}$ in the $x$-axis legend of left panels is the conformal time at recombination ($z\approx 1100$). For $\delta \ln k=0.1$ and $0.03$ the results are compared to CAMB outputs. For $\delta\ln k=0.01$ a simple modification of CAMB cannot be applied as too much memory is required to store the transfer functions for all ($\ell$, $k$) pairs. \label{fig:pkcls}}
\end{figure}

We compute the CMB temperature power spectrum using the enhanced CMB integrator in CosmoLib and compare the results to the smooth-$\primsca(k)$ case. The relative difference between the non-smooth (for a series of $\delta\ln k$) and the smooth model is shown in Figure~\ref{fig:pkcls}. For $\delta \ln k =0.1$ and $\delta \ln k = 0.03$ we compare the results to CAMB output (both with lensing) and find good agreement. The CAMB outputs are obtained by a straightforward modification of CAMB, i.e., increasing the $\ell$ sampling frequency in the input file and increasing the $k$ sampling frequency in the source code. For $\delta \ln k \lesssim 0.01$ the simple modification of CAMB fails due to insufficient memory to store the transfer functions.

For $\delta\ln k\ll 1$, the amplitude of oscillations in the CMB angular power spectrum (right-hand panels) is smaller than that in $\primsca(k)$ (left-hand panels). This suppression is generic when a 3D power spectrum is projected to a 2D one, even though in the CMB case it is further complicated by the finite duration of recombination and the recombination physics \cite{Adshead/etal:2011}. As shown in \cite{Huang/etal:2012}, when the frequency of oscillations is constant in $\ln k$, such as in eq.~\eqref{eq:monoPs}, the relative suppression is given by $\sim \sqrt{\delta\ln k}$, as confirmed by the examples shown in Figure~\ref{fig:pkcls}. Moreover, for $\delta\ln k \lesssim 0.01$, in addition to the projection effect, CMB lensing also significantly smears out the oscillations in $C_\ell$ at high $\ell \gtrsim 2000$. While for $\delta\ln k = 0.1$, the lensing smearing effect is almost negligible. See~\cite{Adshead/etal:2011, Lewis/Challinor:2006} for more detailed discussions about the lensing smearing effect. Finally, note that, although the oscillations in $C_\ell$ are damped, they maintain the same relative width of those of the left-hand panels, i.e., $\delta\ln \ell = \delta\ln k$ where $\ell \gtrsim (\delta\ln k)^{-1}$. At low $\ell$ where $\ell \lesssim 1/\delta\ln k$ the oscillations in $k$ space disappear in $\ell$ space due to the discreteness of $\ell$.

This discussion shows that the enhanced CMB integrator can accurately compute the oscillations in CMB to $\Delta C_\ell/C_\ell \lesssim 10^{-3}$. This does not mean, however, that the total $C_\ell$ is accurate to $10^{-3}$. The $C_\ell$ power spectrum can be systematically biased at subpercent level due to, e.g., recombination uncertainties \cite{Rubino-Mart/etal:2010}. Understanding and eliminating these theoretical errors is important if we want to extract generic features in $\primsca(k)$ with $10^{-3}$ accuracy. On the other hand, if we are only interested in a model predicting a specific feature in $C_\ell$ that cannot be mimicked by other effects, we can focus only on the relative difference in $C_\ell$.

\section{The forecast techniques \label{sec:forecast}}

CosmoLib uses Fisher matrix analysis and MCMC method to forecast the constraints on cosmological parameters for future CMB, LSS and SN experiments. In this section we discuss the modeling of the likelihoods and the parameter estimation methods.

\subsection{The likelihoods}

\subsubsection{CMB simulation}

Given a likelihood function ${\cal L}$, we define $\chi^2 \equiv -2 \ln {\cal L}$. For a nearly full-sky CMB experiment $\chi^2$ can be approximated by \cite{Verde/etal:2006,Baumann/etal:2009}
{\small
\begin{align}
\label{chisq_CMB}
\chi^2 =& \sum_{\ell=\ell_{\rm min}}^{\ell_{\rm max}} (2\ell+1)f_{\rm sky}\, \left[ \frac{\hat{{\cal C}}_\ell^{BB}}{{\cal C}_\ell^{BB}} - 3 + \ln\left(\frac{{\cal C}_\ell^{BB}}{\hat{{\cal C}}_\ell^{BB}}\right)  + \frac{\hat{{\cal C}}_\ell^{TT}{\cal C}_\ell^{EE} + \hat{{\cal C}}_\ell^{EE}{\cal C}_\ell^{TT} - 2\hat{{\cal C}}_\ell^{TE}{\cal C}_\ell^{TE}}{{\cal C}_\ell^{TT}{\cal C}_\ell^{EE}-({\cal C}_\ell^{TE})^2} + \ln{\left(\frac{{\cal C}_\ell^{TT}{\cal C}_\ell^{EE}-({\cal C}_\ell^{TE})^2}{\hat{{\cal C}}_\ell^{TT}\hat{{\cal C}}_\ell^{EE}-(\hat{{\cal C}}_\ell^{TE})^2}\right)}\right] \ ,  
\end{align}}
where $\ell_{\rm min}$ and $\ell_{\rm max}$ are suitable cutoffs that are determined by the observed fraction of sky $f_{\rm sky}$ and the beam resolution of the experiment. In this formula, ${\cal C}^{XY}_\ell$  are the model-dependent theoretical angular power spectra (including the noise contributions) for the temperature, $E$ and $B$ polarizations and their cross-correlations, with $X,Y=\{T,E,B \}$. We compute the noise contribution $N_\ell$ assuming Gaussian beams. The mock data $\hat{{\cal C}}^{XY}_\ell$ are ${\cal C}^{XY}_\ell$  for the fiducial model.

We use the model introduced in \cite{Verde/etal:2006} (and later updated in \cite{Baumann/etal:2009}) to propagate the effect of polarization foreground residuals into the estimated uncertainties on the cosmological parameters. For simplicity, in our simulation we consider only the dominant components in the frequency bands that we are using, i.e., the synchrotron and dust signals.  We assume that foreground subtraction can be done correctly down to a level of 5\%. (This parameter is adjustable by the user.)

\subsubsection{SN simulation}

For the SN simulation, we use the model given by the Dark Energy Task Force (DETF) forecast \cite{Albrecht/etal:2006}. In this case
\begin{equation}
 \chi^2= \sum_i\left(\frac{m_i - \hat{m}_i}{\delta m_i}\right)^2\; ,
\end{equation}
with $i$ going over the SN samples. More specifically, here $m_i$ and $\hat{m}_i$ are the theoretical expectation and observed magnitude of the $i$-th supernova, respectively. The uncertainty $\delta m_i$ is computed by quadratically adding a peculiar velocity (a user-defined constant) to the intrinsic uncertainty in the supernova absolute magnitude (another user-specified constant). 

The apparent magnitude of SN is modeled as
\begin{align}
m &= M -\mu^Lz-\mu^Qz^2 +5\log_{10}\left(\frac{d_L}{\text{Mpc}}\right)+25   - \mu^S\delta_{\text{near}}\ . \label{snm}
\end{align}
The first three terms model the redshift evolution of the absolute magnitude of the supernova peak luminosity. In particular, $M$ is a free parameter with a flat prior over $-\infty<M<+\infty$; for $\mu^L$ and $\mu^Q$, Gaussian priors are applied. The widths of the Gaussian priors are user-input parameters. Finally, given that the nearby samples are likely to be a collection from many other experiments, an offset $-\mu^S\delta_{\text{near}}$, where $\delta_{\text{near}}$ is unity for the nearby samples ($z<z_{\rm near}$) and zero otherwise, is added to account for the systematics. For $\mu^S$ we also apply a Gaussian prior with a user-specified width. The threshold redshift $z_{\rm near}$  is also user-defined. In conclusion, in this model there are four nuisance parameters $(M, \mu^L, \mu^Q, \mu^S)$, which we marginalized analytically. 

\subsubsection{LSS likelihood}

We model the galaxy power spectrum in redshift space as (e.g., \cite{Kaiser:1987, Peacock:1992, Peacock/Dodds:1994})
\begin{equation}
P_g(k,\mu; z) = \left(b+ f \mu^2\right)^2 D^2(z)P_m(k) \exp\left(-k^2\mu^2\sigma_r^2 \right), \label{eq:Pg}
\end{equation}
where $\mu$ is the cosine of the angle between the wave vector $\mathbf{k}$ and the line of sight, $D(z)$ is the linear growth factor, $f \equiv {d\ln D}/{d\ln a}$ is the linear growth rate, $P_m(k)$ is the matter power spectrum today (at $z=0$) and $\sigma_r$ parameterizes the effect of small scales velocity dispersion and redshift errors as explained below. The matter power spectrum $P_m(k)$ is computed using Poisson's equation, that is, $P_m(k) = 4k^4|\Phi_k|^2/(9H^4\Omega^2_m)$.

The term $f \mu^2$  accounts for the redshift distortions due to the large-scale peculiar velocity field \cite{Kaiser:1987}, which is correlated with the matter density field. 
The exponential factor on the right-hand side  accounts for the radial smearing due to the redshift distortions that are uncorrelated with the LSS. In particular, we consider two contributions. The first is due to the redshift uncertainty of the spectroscopic galaxy samples which is estimated to be $\sigma_z=0.001(1+z)$  \cite{Laureijs:2009}. (In CosmoLib the user is allowed to change this value.) The second  comes from the Doppler shift due to the virialized motion of galaxies within clusters, which  typically has  a pairwise velocity dispersion $\sigma_g$ of the order of few hundred $\text{km/s}$. This can be parameterized as $\frac{\sigma_g}{\sqrt{2}} (1+z)$ \cite{Peacock/Dodds:1994}.
The two contributions are quadratically added together 
\begin{equation}
\sigma_r^2 = \frac{(1+z)^2}{H^2(z)}  \left(\sigma_z^2 + {\sigma_g^2}/{2} \right)\;, \label{eq:sigma2}
\end{equation}
where $H(z)$ is the Hubble parameter. 

Practically, neither the redshift measurement nor the virialized motion of galaxies can be precisely modeled. In particular, the radial smearing due to peculiar velocity is not necessarily close to Gaussian. Thus, eq.~(\ref{eq:Pg}) should not be used for wavenumbers $k>\frac{H(z)}{\sigma_g (1+z)}$, where the radial smearing effect is important. We introduce a UV cutoff $k_{\max}$ as the smallest value between $\frac{H}{\sigma_g(1+z)}$ and $\frac{\pi}{2R}$, where $R$ is chosen such that the r.m.s.~linear density fluctuation of the matter field in a sphere with radius $R$ is $0.5$. 

The survey volume is split into $n_z$ redshift bins from $z_{\min}$ to $z_{\max}$, with all these parameters to be specified by the user.  The number density of galaxies that can be used is $\bar{n}=\varepsilon \bar{n}_{\rm obs}$, where $\varepsilon$ is the fraction of galaxies with measured redshift to be specified by the user. Due to the high accuracy of the spectroscopic redshift and the width of the bins, we ignore the bin-to-bin correlations and write $\chi^2$ as 
\begin{equation}
  \chi^2 = \sum_{k,\mu,z\ \rm bins} \left(\frac{P_{g, \rm model} - P_{g, \rm fiducial}}{\Delta P_{g,\rm fiducial}}\right)^2\ .
\end{equation}
As on large scales the matter density field has, to a very good approximation, Gaussian statistics and uncorrelated Fourier modes, the band-power uncertainty is given by \cite{Tegmark/etal:1998}
\begin{equation}
\Delta  P_g = \left[ \frac{2 (2\pi)^3}{(2\pi k^2dk d\mu) (4\pi r^2f_{\rm sky} dr)}\right]^{1/2}\left(P_g+\frac{1}{\bar{n}}\right), \label{eq:dPg}
\end{equation}
where  $r$ is the comoving distance given, for a flat FRW universe, by $r(z)=\int_0^z cdz'/H(z')$. The second term in the parenthesis is due to shot noise, under the assumption that the positions of the observed galaxies are generated by a random Poisson point process. In practice $\bar{n}$ is not known a priori and is calibrated by galaxies themselves. The imperfect knowledge of $\bar{n}$ can bias $P_g$ on the scale of the survey \cite{Tegmark/etal:1998}. This is taken into account by using an IR cutoff $k_{\min}\sim {\rm Gpc}^{-1}$. This is chosen such that $k_{\min}^{(i)} =  2\pi/V^{1/3}_i$, where $V_i$ is the comoving volume of the $i$-th ($i = 1,  \ldots, n_z$) redshift slice. Finally, the user has to specify the binning scheme for $k$ and $\mu$. For $k$ we allow uniform binning in $\ln k$ or in $k$. For $\mu$ only uniform binning in $\mu$ is allowed.

In the special case where $P_m(k)$ has sharp features, we must consider the smearing effect due to the fact that we are only observing a finite volume. This effect is approximated by replacing $P_m(k)$ in \eqref{eq:Pg} with its convolution with a Gaussian window, where the width of the Gaussian window $\sigma_W$ has been chosen to be 
\begin{equation}
\sigma_W = \frac{\sqrt{2 \ln 2}}{2 \pi} \left(\frac{4 \pi}{3} \right)^{1/3} k_{\rm min} \simeq 0.302 \; k_{\rm min}\;. 
\end{equation}
In such a way, the real-space representation of the window, if cut off at its half-height, contains the same volume  as that of the redshift bin. The fact that  $\sigma_W$ is smaller than $k_{\rm min}$ allows us to neglect the overlap between window functions centered around   neighboring values of $k$.

\subsection{Parameter Estimation}

\subsubsection{Fisher Matrix Analysis}

In general, the likelihood can be written as
\begin{equation}
\ln \mathcal{L} (\mathbf{p}; \mathbf{p}_{\rm fid}) = -\frac{1}{2} \left[\mathbf{d}(\mathbf{p}) - \mathbf{d}(\mathbf{p}_{\rm fid})\right]^T C^{-1}(\mathbf{p}; \mathbf{p}_{\rm fid})  \left[\mathbf{d}(\mathbf{p}) - \mathbf{d}(\mathbf{p}_{\rm fid})\right]\;,
\end{equation}
where $\mathbf{d}$ is the data vector, $\mathbf{p}_{\rm fid}$ the fiducial parameter vector,  $\mathbf{p}$ the parameter vector for which one wants to evaluate the likelihood, and $C^{-1}(\mathbf{p}; \mathbf{p}_{\rm fid})$ the covariance matrix.

The fisher matrix for $p_i$, $p_j$ (two components of $\mathbf{p}$) is then
\begin{equation}
F_{ij} \equiv -\left.\frac{\partial^2 \ln \mathcal{L}}{\partial p_i\partial p_j} \right\vert_{\mathbf{p} =\mathbf{p}_{\rm fid}} =\left. \frac{\partial \mathbf{d}(\mathbf{p})}{\partial p_i}  C^{-1}(\mathbf{p}; \mathbf{p}_{\rm fid})  \frac{\partial \mathbf{d}(\mathbf{p})}{\partial p_j} \right\vert_{\mathbf{p} =\mathbf{p}_{\rm fid}}\;,
\end{equation}
where the partial derivatives $ \frac{\partial \mathbf{d}(\mathbf{p})}{\partial p_i}$ can be evaluated numerically:
\begin{equation}
\frac{\partial \mathbf{d}(\mathbf{p})}{\partial p_i} = \frac{1}{2\Delta p_i} \left[\mathbf{d}(p_1, p_2, \ldots,p_i+\Delta p_i, \ldots, p_n)- \mathbf{d}(p_1, p_2, \ldots, p_i-\Delta p_i, \ldots, p_n)\right]\;.
\end{equation} 
The stepsize $\Delta p_i$ is initially supplied by the user, and then adjusted by the software in such a way that the variation of $\chi^2$ is of $O(1)$ when $p_i$ is varied by $\Delta p_i$. By doing this, we have assumed that the likelihood is approximately Gaussian in the proximity of $\mathbf{p}_{\rm fid}$ where the variation of $\chi^2$ is $\lesssim O(1)$. If the likelihood is highly non-Gaussian, Fisher matrix analysis does not give reliable estimations of the error bars of parameters. In this case, one should use the MCMC method to fully explore the structure of the likelihood.

\subsubsection{MCMC method}

CosmoLib has an independent MCMC engine using the Metropolis-Hastings algorithm. The traditional approach is to define the proposal density $Q(\vecx; \vecx')$ (the probability of walking from $\vecx$ to $\vecx '$ in the parameter space) using a roughly estimated covariance matrix $C_e$
\begin{equation}
Q(\vecx; \vecx') \propto \exp{\left[-\frac{1}{2}(\vecx-\vecx')^TC_e^{-1} (\vecx-\vecx')\right]} \;. \label{eq:prop}
\end{equation}
Convergence can be achieved quickly if $C_e$ is close the posterior covariance matrix of $\vecx$.

However, sometimes we need to treat models where the likelihood periodically depends on some phase parameters. Here we take the axion monodromy inflation model for example. The likelihood $\mathcal{L}$ is a periodic function of the axion phase parameter $\varphi$,
\begin{equation}
\mathcal{L}(P, \varphi) = \mathcal{L}(P, \varphi + 2\pi)\;,
\end{equation}
where we have used $P$ to represent the collection of other parameters. If $\varphi$ is not well constrained, we will obtain multi-branches in the posterior, i.e., for a fixed value of $\varphi$ and a chosen confidence level, the allowed values of $P$ locate in well separated regions in the parameter space.

Intuitively the separated regions can be more efficiently explored by restricting the range  of $\varphi$ to one period and proposing with wrap-around or, in a more rigorous language, by using a periodic proposal density. For $\vecx = (P, \varphi)$ and $\vecx' = (P', \varphi')$, we use
\begin{equation}
Q(P, \varphi;P',\varphi' )\propto  \sum_{n=-\infty}^\infty\exp{\left[-\frac{1}{2}(\vecx-\vecx'_n)^TC_e^{-1} (\vecx-\vecx'_n)\right]} \; , \label{eq:propupdate}
\end{equation}
where $x'_n \equiv (P', \varphi'+2n\pi)$. The estimation of covariance matrix, $C_e$, is practically computed with a trial run that is terminated before the multi-branches of the posterior are explored by the random walk.

We find that the periodic proposal density \eqref{eq:propupdate} leads to significant improvement of the convergence. For the axion monodromy model, it takes about 5-10 times longer to achieve convergence using \eqref{eq:prop} than using \eqref{eq:propupdate}.

The output chains in CosmoLib have the same format as those in CosmoMC~\cite{Lewis/Bridle:2002}. The chains can hence be directly analyzed using the GetDist tool in CosmoMC. For completeness, an independent postprocessing tool is supplied in CosmoLib to analyze and visualize the marginalized posterior of parameters. In the online documentation the reader can find the instructions on how to use this tool.

\subsubsection{Oscillations in the Current CMB Data?}

Recently a hint of the axion monodromy cosine oscillations (see eq.~\eqref{eq:monoPs}) in WMAP-7yr \cite{Komatsu/etal:2011, Larson/etal:2011} and ACT CMB data \cite{Dunkley/etal:2011} has been claimed in Ref.~\cite{Aich/etal:2011}.  Ref.~\cite{Meerburg/etal:2012} confirms the finding that $\chi^2$ can be significantly improved in some regions of parameter space where oscillations in the primordial power spectrum are assumed. In this section we use CosmoLib to constrain the axion monodromy model with the same data sets. We find that when the CMB power spectrum is accurately computed and rigorous statistical method is used, there is {\it no} detectable axion monodromy oscillations in the CMB data.

\begin{figure}
\includegraphics[width=\halffigurewidth]{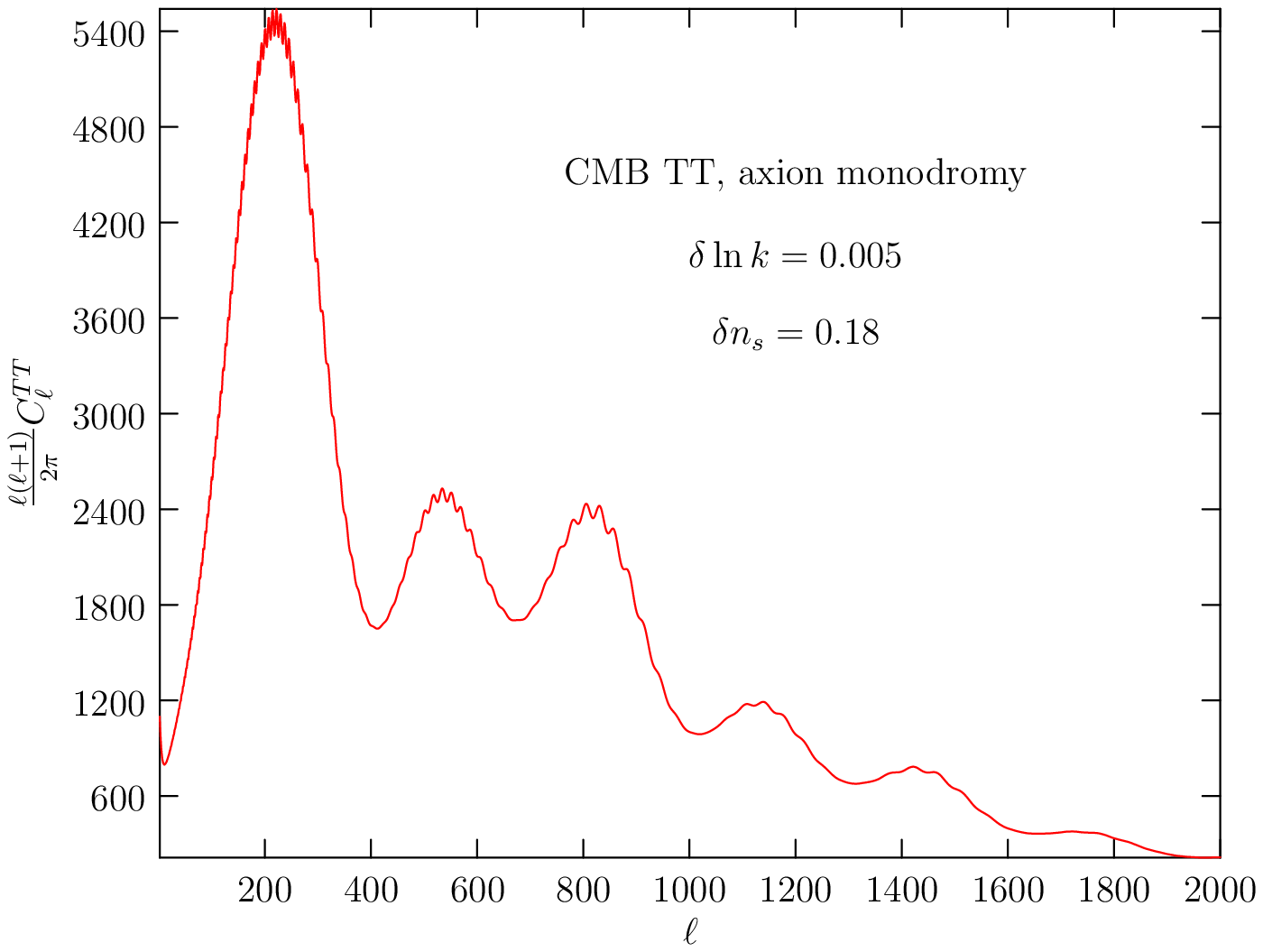}
\caption{The CMB angular power spectrum for axion monodromy model with $\delta\ln k=0.005$, $\delta n_s = 0.18$. The other cosmological parameters are $\Omega_{b0}h^2 = 0.0223$, $\Omega_{c0}h^2 = 0.1119$, $\theta = 1.041$, $\tau_{\rm re} = 0.0884$, $n_s = 0.975$, $\ln (10^{10}A_s) = 3.04$. Modulation of $C_\ell$ is uniform in $\ln \ell$ and is almost invisible at high-$\ell$ due to lensing smearing. This should be compared to Fig.~5 in Ref.~\cite{Aich/etal:2011}, where the random fluctuations in $C_\ell$ implies insufficient numerical accuracy of the modified CAMB used therein. \label{fig:am_Cls}}
\end{figure}

In Refs.~\cite{Aich/etal:2011} the authors used their modified CAMB to compute the CMB power spectrum. As discussed in previous sections, such a modification is not trivial for $\delta\ln k \lesssim 10^{-2}$. Since the best-fit $\delta\ln k$ found in Ref.~\cite{Aich/etal:2011} is small -- $\delta\ln k \approx 0.005$ (derived from Table III of Ref.~\cite{Aich/etal:2011} and equation 51 in Ref.~\cite{Huang/etal:2012}), it is necessary to exam the numerical accuracy of the modified CAMB used in~\cite{Aich/etal:2011}. For $\delta\ln k\approx 0.005$, the modulation period in $\ln k$ is $T_{\ln k} = 2\pi \delta\ln k\approx 0.03$. In the CMB power spectrum one should see same modulation period in $\ln \ell$, i.e., $T_{\ln \ell} = T_{\ln k} \approx 0.03$. Thus, from $\ell = 1000$ to $\ell  = 1200$ there should be about $7$ oscillations in $C_\ell$. However, in Figure 5 of Ref.~\cite{Aich/etal:2011} the number of oscillations in $C_\ell$ between $\ell = 1000$ and $\ell = 1200$ are much more than $7$. This implies that the ``modulations'' in $C_\ell$  shown in Ref.~\cite{Aich/etal:2011} may just be numerical noises. In Figure~\ref{fig:am_Cls} we show the CMB temperature angular power spectrum computed with CosmoLib, where the parameters are chosen to be close to the ones used in Figure 5 of Ref.~\cite{Aich/etal:2011}. Qualitative difference can be seen between the two figures. The $C_\ell$ spectrum computed using CosmoLib presents clear modulations that agrees with the $\delta\ln k$ value, while the modified CAMB used in Ref.~\cite{Aich/etal:2011} failed to produce the expected modulations.

\begin{figure}
\includegraphics[width=\halffigurewidth]{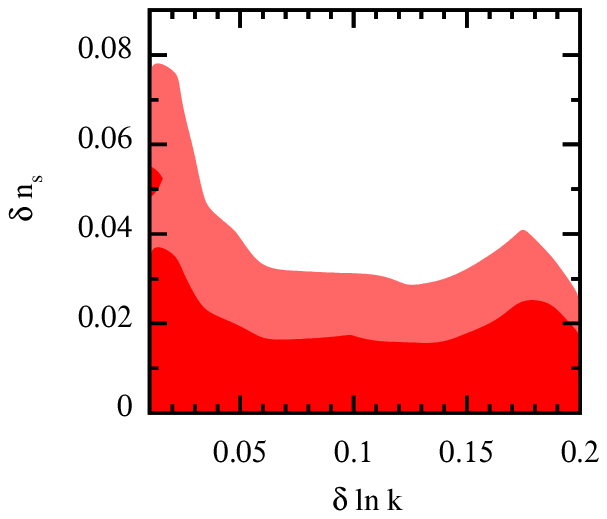}
\caption{The marginalized 68.3\% and 95.4\% confidence level contours of $\delta n_s$ and $\delta\ln k$ for axion monodromy model. WMAP-7yr and ACT data are used. CMB angular power spectrum are computed up to $\ell = 4000$ with CMB lensing effect included. Uniform priors $0.003\le \delta\ln k\le 0.2$ and $0\le n_s\le 0.2$ are used.  No detection of axion monodromy oscillations are found since zero amplitude of oscillations ($\delta n_s = 0$) is consistent with the data. \label{fig:deltansdeltalnk}}
\end{figure}

In Ref.~\cite{Huang/etal:2012} we pointed out that, a significant improvement of $\chi^2$ does not necessarily imply a detection of models with periodic oscillations, which typically has a spiky likelihood that is highly non-Gaussian. A rigorous treatment is to compute the marginalized probability of the amplitude of oscillations $\delta n_s$. The marginalization should be done in such a way that all the other cosmological and nuisance parameters are allowed to vary. A detection of monodromy oscillations should not be claimed unless $\delta n_s=0$ is excluded by the data. We did the full marginalization using MCMC method. The CMB power spectra are computed using the accurate integrator in CosmoLib. The marginalized 68.3\% and 95.4\% confidence level posterior contours are shown in Figure~\ref{fig:deltansdeltalnk}.

\section{Conclusions \label{sec:conclusion}}

We introduced the numerical package CosmoLib and focused on its features that are complementary to other numerical codes. The major advantage of CosmoLib is that it can accurately compute CMB angular power spectrum for inflationary models that predict sharp features in the primordial power spectrum of metric perturbations. This is not available in any other publicly available CMB codes. 

CosmoLib can calculate the relative fluctuations in $C_\ell$ to accuracy $\sim 10^{-3}$. Because of cosmic variance, we cannot measure $C_\ell$ to this accuracy {\it if all  $C_\ell$ are treated independently}. However, our purpose is to use CosmoLib to study specific models where the degrees of freedom in the $C_\ell$ spectrum is small. In other words, if we assume a specific model (such as the axion monodromy model), the relative error in $C_\ell$ can be constrained to a level that is well below cosmic variance. In the naive limit where the $C_\ell$ spectrum is controlled by a single scaling parameter $s$, that is, $C_\ell = s C_{\ell, \rm fiducial}$, we can constrain $C_\ell$ to a relative accuracy $1/\sqrt{N} \approx 1/\ell_{\max}$, where $N= \sum_\ell (2\ell +1) \approx \ell_{\max}^2$ is the total number of measured spherical harmonic modes. For a future experiment that measures $C_\ell$ to cosmic variance for $\ell$ up to a few thousands \cite{PlanckScienceTeam:2009}, the aforementioned $10^{-3}$ relative accuracy is necessary.

While a straightforward  (but not optimized) modification of CAMB and CLASS to use non-smooth $P(k)$ seems to be trivial, in practice it is often limited by the available memory and tolerable computation time. We pointed out that the modified CAMB in Ref.~\cite{Aich/etal:2011} produces numerical noises instead of the expected modulation in $C_\ell$ spectrum. Repeating the computation in Ref.~\cite{Aich/etal:2011} using CosmoLib and the same data sets (WMAP + ACT), we found no detection or hint of axion monodromy model in the current CMB data.

This forecast toolkit contains a fisher matrix calculator, a MCMC engine, a postprocessing tool for chain analysis, and likelihoods for future CMB, galaxy survey, and supernova observations. The MCMC engine has an option of using a periodic proposal density, which can significantly accelerate the convergence of the chains in the case where the likelihood is a periodic function of some parameters. Although the likelihood models in CosmoLib are likely to be too simple for  real experiments with complicated specifications, they provide a quick {\it estimation} of the performance of future CMB/LSS/SN experiments, for which the details of specifications are not yet known. We are planning to include more likelihoods for, e.g., weak lensing experiments in future releases. 

\acknowledgements 
I thank Licia Verde, Filippo Vernizzi, Cyril Pitrou,  Julien Lesgourgues and Emiliano Sefusatti for useful advice and discussions.


\end{document}